\documentclass[aps,prl,showpacs,twocolumn]{revtex4}
\usepackage{amsmath, amssymb}
\usepackage{graphicx}
\usepackage{subfigure}

\renewcommand{\vec}[1]{\mathbf{#1}}

\newcommand{\cici}{{\rm\raisebox{-2.pt}{$\>\stackrel{\scriptstyle\circ}{\scriptstyle\circ}\>$}}}

\begin{document}

\title{Universal measurement of quantum correlations of radiation}

\author{E. Shchukin}
\email{evgeny.shchukin@uni-rostock.de}
\author{W. Vogel}
\email{werner.vogel@uni-rostock.de} \affiliation{Arbeitsgruppe Quantenoptik, Fachbereich Physik,
Universit\"at Rostock, D-18051 Rostock, Germany}

\begin{abstract}
  A measurement technique is proposed which, in principle, allows one to
  observe the general space-time correlation properties of a quantized
  radiation field. Our method, called balanced homodyne correlation
  measurement, unifies the advantages of balanced homodyne detection with
  those of homodyne correlation measurements.
\end{abstract}

\pacs{03.65.Wj, 42.50.Ar, 42.50.Dv}

\maketitle

The pioneering photon correlation experiments performed by Hanbury Brown and
Twiss~\cite{n-177-27} in the 1950th have stimulated a series of investigations
of the correlation properties of radiation fields. The quantum field
theoretical description of the coherence properties of radiation was
introduced by Glauber~\cite{pr-130-2529}. In this approach the considered
correlation functions contain equal powers of negative and positive frequency
parts of field operators, which is closely related to the possibilities of
observation by photon correlation measurements.

A complete description of the quantum statistical properties of radiation also
requires the consideration of more general space-time dependent correlation
functions~\cite{rmp-37-231, kl-sud, perina, mandel-wolf}, which are composed
of unequal powers of photon annihilation and creation operators.
Phase-sensitive correlations of such a type are not directly accessible by
photoelectric detection. In principle, they could be observed by transmitting
the radiation, prior to the measurement by photodetectors, through an
appropriately chosen nonlinear medium~\cite{mandel-wolf}. The realization of
such measurements is difficult, in particular, when microscopic fields to be
measured are too weak for causing the needed nonlinear effects.

Usually phase-sensitive radiation properties are measured by homodyne
detection~\cite{ieee-tit-24-657}, where the fields to be measured are
superimposed with a coherent reference field, the so-called local oscillator.
The methods of balanced homodyne tomography~\cite{prl-70-1244} and of
unbalanced homodyning~\cite{pra-53-4528} allow one to reconstruct the quantum
state of an effective single-mode radiation field. The reconstruction of
moments has also been considered \cite{pra-53-1197}. More complex methods of
multiport homodyning allow one to observe space-time dependent
correlations~\cite{pra-51-4240}. However, involved reconstruction methods are
needed and one obtains only insight in smoothed quantum states, due to noise
effects caused by imperfect detection, for details see~\cite{welsch99} and
references therein.

Some special radiation properties composed of unequal orders of annihilation
and creation operators can be measured by homodyne correlation
techniques~\cite{prl-67-2450}. In this case imperfect detection does not
contaminate the detected correlation functions. The method has been further
developed for the measurement of arbitrary moments of a single-mode radiation
field~\cite{pra-72-043808}, with the aim to characterize the nonclassical
properties of radiation.  However, the use of a weak local oscillator
makes it difficult to determine moments of
high orders.

Why is the accurate determination of space-time dependent correlation
functions of radiation fields, including those composed of unequal powers of
annihilation and creation operators, of interest? It would lead to new
possibilities to study the general (high-order) quantum coherence properties
of radiation sources, including the dynamical properties and the spatial
irradiation characteristics. New types of time-dependent nonclassical
correlation properties could be investigated, which generalize known effects
like photon antibunching~\cite{prl-39-691}. Last but not least, the
characterization of entanglement of continuous quantum states can been based
on such correlation functions~\cite{prl-95-230502}. Hence their measurement
is of great interest for any kind of application of nonclassical
and, in particular, of entangled radiation fields.

In this letter we propose a method for observing the most general normally-
and time-ordered correlation functions of radiation fields. It combines
advantages of balanced homodyning with those of homodyne correlation
measurements in a method to be called balanced homodyne correlation (BHC)
measurement. A chosen correlation function is determined by a fixed number of
photodetectors, so that imperfect detection does not lead to smoothing
effects.  Since a strong local oscillator can be used in the BHC method, the
signal-to-noise ratio allows to determine high-order correlation functions.

Let us consider the general correlation function
$\mathcal{G}^{(n,m)}(x_1, \ldots, x_n, y_1, \ldots, y_m)$ of an
electromagnetic field,
\begin{equation}\label{eq:G}
\begin{split}
    \mathcal{G}^{(n,m)}(x_1, &\ldots, x_n, y_1, \ldots, y_m) = \\
    &\left\langle \cici \prod^n_{k=1}\hat{\mathcal{E}}^{(-)}(x_k)
    \prod^m_{l=1}\hat{\mathcal{E}}^{(+)}(y_l) \cici \right\rangle,
\end{split}
\end{equation}
where $x_k = (\vec{r}_k, t_k)$, $y_l = (\vec{s}_l, \tau_l)$ refer to both
space and time points. The operator $\hat{\mathcal{E}}^{(-)}$ ($\hat{\mathcal{E}}^{(+)}$) denotes the negative (positive) frequency part of the electric field operator containing the photon creation (annihilation) operators. The notation $\cici \cici$, as used in \cite{book},
means that field operators are to be written in normal order ($\hat{\mathcal{E}}^{(-)}$ to the left of $\hat{\mathcal{E}}^{(+)}$), and time order (time
arguments increasing to the right in products of $\hat{\mathcal{E}}^{(-)}$ and to the
left in products of $\hat{\mathcal{E}}^{(+)}$). For simplicity we restrict ourselves to the case of one polarization, the extension to different polarizations is straightforward.

\begin{figure}
    \includegraphics{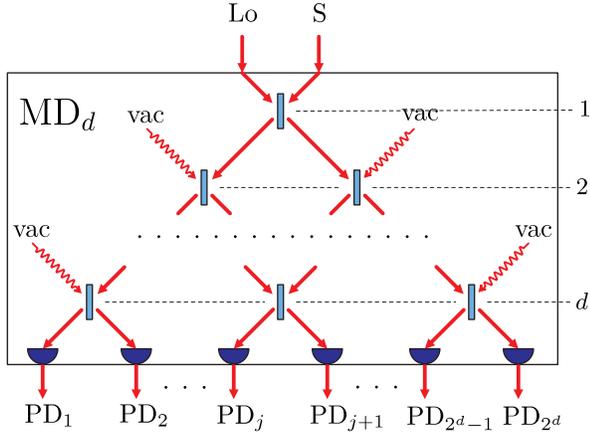}
\caption{Basic measurement device
$\mathrm{MD}_d$.}\label{fig:scheme}
\end{figure}

\begin{figure}
    \includegraphics{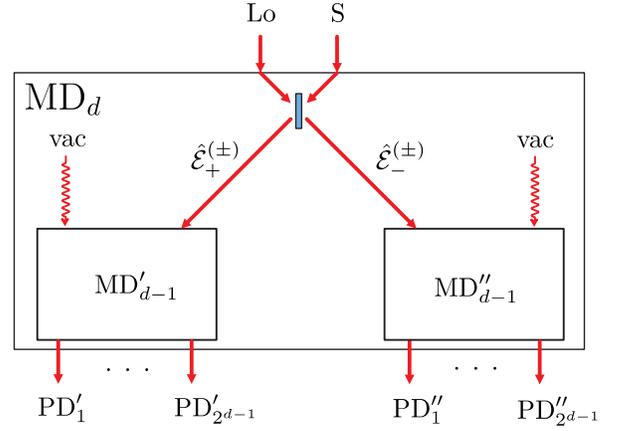}
\caption{The recursive structure of
$\mathrm{MD}_d$.}\label{fig:scheme2}
\end{figure}

Let us start with the simplest case of one and the same space-time point in all the field operators in the expression \eqref{eq:G}: $x_1 = \ldots = x_n = y_1 = \ldots =
y_m = (\vec{r}, t)$. In such a case the correlation function \eqref{eq:G} reads as
\begin{equation}\label{eq:Gnm}
    \mathcal{G}^{(n,m)}(\vec{r}, t) =
    \left\langle\hat{\mathcal{E}}^{(-)}(\vec{r}, t)^n
    \hat{\mathcal{E}}^{(+)}(\vec{r}, t)^m\right\rangle.
\end{equation}
Here the field operators are already normally ordered and time-ordering
becomes meaningless. The correlation function \eqref{eq:Gnm} can be measured
with the device shown in Fig. \ref{fig:scheme}. This device is parameterized
by an integer, the number of levels or depth of the device. The measurement
device ($\mathrm{MD}$) of the depth $d$ we denote by $\mathrm{MD}_d$. It has
$2^d$ photodetectors and can be composed recursively of
two devices $\mathrm{MD}^{\prime}_{d-1}$ and
$\mathrm{MD}^{\prime\prime}_{d-1}$ of lower depth $d-1$, cf. Fig.
\ref{fig:scheme2}. The elementary building block of our setup is the lowest
order device $\mathrm{MD}_1$. Roughly speaking, $\mathrm{MD}_d$ can be
constructed by replacing both photodetectors of $\mathrm{MD}_1$ by
$\mathrm{MD}_{d-1}$. All the beamsplitters of the device $\mathrm{MD}_d$ are
assumed to be symmetric $50$\%-$50$\% and all the photodetectors to have the
same quantum efficiency $\eta$. This can be realized by balancing them with
polarizers.  \nocite{prl-87-253601}

In the BHC approach the device $\mathrm{MD}_d$ allows us to measure the
correlations \eqref{eq:Gnm} with $n+m \leqslant 2^{d-1}$, so that the minimal
depth $d$ necessary to measure the moments \eqref{eq:Gnm} with given $n$ and
$m$ is $d = \lceil \log_2(n+m) \rceil + 1$ ($\lceil x \rceil$ is the smallest
integer number greater or equal to $x$). We assume the local oscillator to be
in a coherent state
\begin{equation}
    \Bigl\langle\hat{\mathcal{E}}^{(-)}_{\mathrm{LO}}\ldots\Bigr\rangle =
    E e^{-i(\omega t + \varphi_{\mathrm{LO}})}, \quad
    \Bigl\langle\ldots\hat{\mathcal{E}}^{(+)}_{\mathrm{LO}}\Bigr\rangle =
    E e^{i(\omega t + \varphi_{\mathrm{LO}})}. 
\end{equation}
One can easily see that the field operators
$\hat{\mathcal{E}}^{(\pm)\prime}_j$
($\hat{\mathcal{E}}^{(\pm)\prime\prime}_j$) detected by $j$-th
photodetector $\mathrm{PD}^\prime_j$
($\mathrm{PD}^{\prime\prime}_j$) of the left subdevice
$\mathrm{MD}^\prime_{d-1}$ (right subdevice
$\mathrm{MD}^{\prime\prime}_{d-1}$) are proportional to the operators
$\hat{\mathcal{E}}^{(\pm)}_+$ ($\hat{\mathcal{E}}^{(\pm)}_-$):
\begin{equation}\label{eq:EE}
    \hat{\mathcal{E}}^{(\pm)\prime}_j = \frac{e^{\pm i \Phi_j}}{\sqrt{2^{d-1}}}\hat{\mathcal{E}}^{(\pm)}_+ + \mathrm{vac}, \quad
    \hat{\mathcal{E}}^{(\pm)\prime\prime}_j = \frac{e^{\pm i \Phi_j}}{\sqrt{2^{d-1}}}\hat{\mathcal{E}}^{(\pm)}_- + \mathrm{vac},
\end{equation}
$j=1, \ldots, 2^{d-1}$, the phase $\Phi_j$ depends
on the path of the signal to the $j$-th photodetector. The operators
$\hat{\mathcal{E}}^{(+)}_\pm$ read as
\begin{equation}
\hat{\mathcal{E}}^{(+)}_\pm = \frac{e^{i \Phi_\pm}}{\sqrt{2}}\Bigl(\hat{\mathcal{E}}^{(+)}(\vec{r}, t)\pm
     i\hat{\mathcal{E}}^{(+)}_{\mathrm{LO}}\Bigr),
\end{equation}
with $\Phi_+-\Phi_- = \pi/2$. The symbol $\mathrm{vac}$ in
\eqref{eq:EE} means vacuum terms which play no role in the considered 
homodyne correlation measurements.

Let us denote by $\Gamma_{j_1, \ldots, j_k}$ the normally-ordered (symbolized
by $::$) correlation function of the photodetectors $\mathrm{PD}_{j_1},
\ldots, \mathrm{PD}_{j_k}$,
\begin{equation}\label{eq:G3}
    \Gamma_{j_1, \ldots, j_k} =
    \Bigl\langle:\hat{\mathcal{E}}^{(-)}_{j_1}
    \hat{\mathcal{E}}^{(+)}_{j_1}\ldots
    \hat{\mathcal{E}}^{(-)}_{j_k}
    \hat{\mathcal{E}}^{(+)}_{j_k}:\Bigr\rangle,
\end{equation}
For any $k=1, \ldots, 2^{d-1}$ and for any $l$, $0 \leqslant l \leqslant k$,
let us select $l$ photodetectors from $\mathrm{MD}'_{d-1}$ and $k-l$ from
$\mathrm{MD}^{\prime\prime}_{d-1}$ (cf. Fig. \ref{fig:scheme2}) and measure
their correlation function. Due to the relations \eqref{eq:EE} such a
correlation function depends only on the numbers $k$ and $l$ but not on the
individual photodetectors chosen. We denote it by $\Gamma^{(k)}_l$, using
Eq.~\eqref{eq:EE} it reads as
\begin{equation}
    \Gamma^{(k)}_{l} = 2^{- k (d-1)}
    \Bigl\langle:\hat{\mathcal{N}}^l_+
    \hat{\mathcal{N}}^{k-l}_-:\Bigr\rangle.
\end{equation}
The operators $\hat{\mathcal{N}}_+$ and $\hat{\mathcal{N}}_-$
are defined analogously to the photon number operator of a single-mode
field,
\begin{equation}\label{eq:N}
    \hat{\mathcal{N}}_\pm = \hat{\mathcal{E}}^{(-)}_\pm
    \hat{\mathcal{E}}^{(+)}_\pm =
    \frac{1}{2}\Bigl(\hat{\mathcal{E}}^{(-)}
    \hat{\mathcal{E}}^{(+)} \pm E \hat{\mathcal{X}}_\varphi + E^2\Bigr),
\end{equation}
and $\hat{\mathcal{X}}_\varphi$ corresponds to
the quadrature operator,
\begin{equation}
    \hat{\mathcal{X}}_\varphi = \hat{\tilde{\mathcal{E}}}^{(+)} e^{-i\varphi} +
    \hat{\tilde{\mathcal{E}}}^{(-)} e^{i \varphi},
\end{equation}
with $\hat{\tilde{\mathcal{E}}}^{(\pm)} = \hat{\mathcal{E}}^{(\pm)}
e^{\pm i \omega t}$ being slowly varying fields and $\varphi =
\varphi_{\mathrm{LO}}+ \pi/2$.

\begin{figure}
    \includegraphics{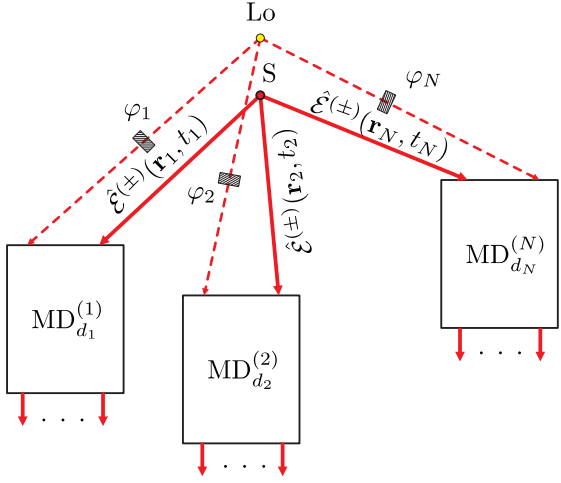}
\caption{Measurement scheme for space-time correlations.}\label{fig:scheme3}
\end{figure}

Let us chose one correlation function $\Gamma^{(k)}_{l}$ for each
$l=0, 1, \ldots, k$ and combine them as
\begin{equation}\label{eq:F}
    F^{(k)} = \sum^k_{l=0}(-1)^{k-l} \binom{k}{l} \Gamma^{(k)}_{l}.
\end{equation}
It is important to note that all the terms in this sum are proportional to $k$-th power of the quantum efficiency of the photodetectors. Using the binomial formula we obtain
\begin{equation}
    F^{(k)}(\varphi) = 2^{-k d}\Bigl\langle:(\hat{\mathcal{N}}_+ - \hat{\mathcal{N}}_-)^k:\Bigr\rangle =
    2^{-k d} E^k \Bigl\langle:\hat{\mathcal{X}}^k_\varphi:\Bigr\rangle,
\end{equation}
which represents the set of BHC data to be analyzed. The dependence of the quantity $F^{(k)}$ on the phase $\varphi$ is given explicitly: $F^{(k)} = F^{(k)}(\varphi)$. According the definition of the operator $\hat{\mathcal{X}}_\varphi$, the normally ordered power $:\hat{\mathcal{X}}^k_\varphi:$ can be expanded as
\begin{equation}
    \langle:\hat{\mathcal{X}}^k_\varphi:\rangle = \sum^k_{l=0} \binom{k}{l}
    \Bigl\langle\hat{\tilde{\mathcal{E}}}^{(-)l} \hat{\tilde{\mathcal{E}}}^{(+)k-l} \Bigr\rangle
    e^{-i (k-2l) \varphi}.
\end{equation}
Hence, the moments $\Bigl\langle\hat{\tilde{\mathcal{E}}}^{(-)n} \hat{\tilde{\mathcal{E}}}^{(+)m}\Bigr\rangle$ can
be obtained directly by Fourier transforming the BHC data:
\begin{equation}
    \Bigl\langle\hat{\tilde{\mathcal{E}}}^{(-)n} \hat{\tilde{\mathcal{E}}}^{(+)m} \Bigr\rangle \sim \int^{2\pi}_0 F^{(n+m)}(\varphi) e^{-i(n-m)\varphi} \, d\varphi.
\end{equation}
The moments of the original field are given by
\begin{equation}
    \Bigl\langle\hat{\mathcal{E}}^{(-)n} \hat{\mathcal{E}}^{(+)m}\Bigr\rangle =
    \Bigl\langle\hat{\tilde{\mathcal{E}}}^{(-)n} \hat{\tilde{\mathcal{E}}}^{(+)m}\Bigr\rangle
    e^{i(n-m)\omega t}.
\end{equation}

Once we know how to measure the moments~\eqref{eq:Gnm} in a single space-time
point, the method can be extended to the general space-time
correlations~\eqref{eq:G}. All field operators in the same space-time points
can be grouped together and after a proper permutation any correlation
function \eqref{eq:G} can be represented in the form
\begin{equation}\label{eq:G2}
    \mathcal{G} = \left\langle \cici \prod^N_{i=1} \hat{\mathcal{E}}^{(-)n_i}(\vec{r}_i, t_i) \hat{\mathcal{E}}^{(+)m_i}(\vec{r}_i, t_i) \cici \right\rangle,
\end{equation}
where some of the $n_i$ or $m_i$ can be zero. For example, the
correlation function
\begin{equation}
    \mathcal{G}^{(1, 2)}(x, y, y) =
    \langle\cici\hat{\mathcal{E}}^{(-)}(x)\hat{\mathcal{E}}^{(+)2}(y)\cici\rangle  
\end{equation}
can be written in the form \eqref{eq:G2} as
\begin{equation}
    \mathcal{G}^{(1, 2)}(x, y, y) =
    \langle\cici\hat{\mathcal{E}}^{(-)1}(x)\hat{\mathcal{E}}^{(+)0}(x)\hat{\mathcal{E}}^{(-)0}(y)\hat{\mathcal{E}}^{(+)2}(y)\cici\rangle. 
\end{equation}

\begin{figure}
    \includegraphics{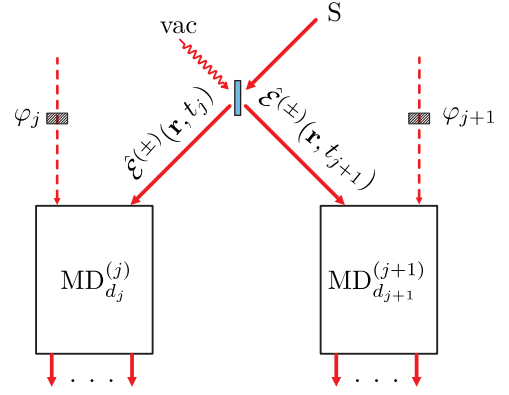}
\caption{Measurement scheme for time-dependent correlations at equal space points, $\vec{r}_j = \vec{r}_{j+1} =
  \vec{r}$.}\label{fig:scheme4}
\end{figure}

Figure \ref{fig:scheme3} illustrates the scheme for measuring the
general correlation function \eqref{eq:G2}. The phases $\varphi_1,
\ldots, \varphi_N$ of the local oscillator in each channel can be
controlled independently by the corresponding phase shifters. For
$\vec{r}_j = \vec{r}_{j+1} = \vec{r}$ and $t_j \not= t_{j+1}$
we split the channel corresponding to the space point
$\vec{r}$ into two parts as shown in Fig. \ref{fig:scheme4}. If more
than two space points coincide, we split the signal in the
corresponding channel by using the same tree-like combination of beam
splitters and photodetectors as in $\mathrm{MD}_d$ for a
proper $d$, see Fig. \ref{fig:scheme}.

Measuring correlations of photodetectors of all the devices
$\mathrm{MD}^{(i)}_{d_i}$, one can extract the functions \eqref{eq:G2} with
$n_i+m_i \leqslant 2^{d_i-1}$, $i=1, \ldots, N$.  Let us select $k_i$
photodetectors from the $i$-th device $\mathrm{MD}^{(i)}_{d_i}$ for $i=1,
\ldots, N$ and measure the correlation function of all the $K=k_1+\ldots+k_N$
photodetectors chosen. One can easily see that such a correlation function
$\Gamma = \Gamma^{(k_1 \ldots k_N)}_{l_1 \ldots l_N}$ has the form
\begin{equation}
    \Gamma^{(k_1 \ldots k_N)}_{l_1 \ldots l_N} = 2^{-P}
    \left\langle \cici \prod^N_{i=1} \hat{\mathcal{N}}^{(i) l_i}_+(\vec{r}_i, t_i) \hat{\mathcal{N}}^{(i) k_i-l_i}_-(\vec{r}_i, t_i) \cici \right\rangle, 
\end{equation}
where $P = \sum_i k_i d_i$ and $l_i$ is the number of those photodetectors of
$\mathrm{MD}^{(i)}_{d_i}$ that belong to the left subdevice
$\mathrm{MD}^{(i)\prime}_{d_i-1}$. Generalizing the single-mode approach, one
can compose the recorded data in the form
\begin{equation}
    F^{(k_1 \ldots k_N)} = \sum^{k_1, \ldots, k_N}_{l_1, \ldots, l_N=0}
    (-1)^{K - L} 
    \binom{k_1}{l_1} \ldots \binom{k_N}{l_N}
    \Gamma^{(k_1 \ldots k_N)}_{l_1 \ldots l_N}, 
\end{equation}
where $L = l_1 + \ldots + l_N$, which is equal to
\begin{equation}
\begin{split}
    F^{(k_1 \ldots k_N)} &= 2^{-P } \left\langle \cici \prod^N_{i=1} (\hat{\mathcal{N}}^{(i)}_+(\vec{r}_i, t_i) - \hat{\mathcal{N}}^{(i)}_-(\vec{r}_i, t_i))^{k_i} \cici \right\rangle \\
    &= 2^{-P} E^K \langle \cici \hat{\mathcal{X}}^{k_1}_{\varphi_1}(\vec{r}_1, t_1) \ldots \hat{\mathcal{X}}^{k_N}_{\varphi_N}(\vec{r}_N, t_N) \cici \rangle. 
\end{split}
\end{equation}
Using multidimensional Fourier transform one can extract from the
function $F^{(k_1 \ldots k_N)} = F^{(k_1 \ldots k_N)}(\varphi_1,
\ldots, \varphi_N)$ the correlations \eqref{eq:G2} or, equivalently,
the original form \eqref{eq:G} of the space-time dependent field
correlation function.

Let us define the characteristic function $C[\vec{u}]$ via
\begin{equation}
    C[\vec{u}] = \left\langle \cici \exp\left(\sum^N_{i=1} (u_i \hat{\mathcal{E}}^{(-)}(x_i) - u^*_i \hat{\mathcal{E}}^{(+)}(x_i))\right) \cici \right\rangle, 
\end{equation}
where $\vec{u} = (u_1, \ldots, u_N)$. The moments \eqref{eq:Gnm} are readily
derived by
\begin{equation}
    \left.\frac{\partial^{n+m} C[\vec{u}]}{\partial^n u_i \partial^m u^*_i}\right|_{\vec{u}=0} = 
    \langle\cici \hat{\mathcal{E}}^{(-)n}(x_i) \hat{\mathcal{E}}^{(+)m}(x_i) \cici\rangle.
\end{equation}
The general correlation functions \eqref{eq:G2} are given by the general
partial derivatives of $C[\vec{u}]$ with respect to all the variables. 
Expanding the characteristic function as
\begin{equation}
    C[\vec{u}] = \sum^{+\infty}_{n_i, m_i = 0} \frac{\langle\cici \prod^N_{i=1}\hat{\mathcal{E}}^{(-)n_i}(x_i) \hat{\mathcal{E}}^{(+)m_i}(x_i) \cici\rangle}{\prod^N_{i=1}n_i!\,m_i!},
\end{equation}
it is seen that, in principle, the correlation functions \eqref{eq:G2}
completely describe the quantum statistical properties of the radiation field
in the chosen space-time points.

Finally, we note that our approach works only for quasimonochromatic fields
that are completely recorded by the detectors. More generally, broadband
fields could be studied by including spectral correlation measurements as was
done for light pulses \cite{prl-87-253601}. This experiment also
shows the feasibility of multichannel measurements as needed for our method.

In conclusion, we have proposed a method for measuring general space-time
dependent correlation functions of quantized radiation fields.  Homodyne
correlation measurements are performed and the data are combined and analyzed
in a balanced form. The detected correlation functions are
insensitive to imperfect detection. By using a strong local oscillator, 
even higher-order correlations can be determined. This opens new
perspectives for the study of nonclassical correlations and, in
particular, of entanglement of complex radiation fields.

The authors acknowledge support by the Deutsche Forschungsgemeinschaft through SFB 652 and GK 567.

\end{document}